\documentclass{aip-cp}
\usepackage[numbers]{natbib}
\usepackage{rotating}
\usepackage{graphicx}

\begin{document}

\title{Investigation of 2$\beta$ Decay of $^{116}$Cd with the Help of Enriched $^{116}$CdWO$_4$ Crystal Scintillators}

\author[kinr]{O.G.~Polischuk\corref{cor1}}
\author[itep]{A.S.~Barabash}
\author[infn-roma2,univ-roma2]{P.~Belli}
\author[infn-roma2,univ-roma2]{R.~Bernabei}
\author[infn-roma1]{F.~Cappella}
\author[lngs]{V.~Caracciolo}
\author[infn-roma2,univ-roma2]{R.~Cerulli}
\author[kinr,kavli]{D.M.~Chernyak}
\author[kinr,csnsm]{F.A.~Danevich}
\author[infn-roma2,univ-roma2]{S.~d'Angelo}
\eaddress{Deceased}
\author[infn-roma1,univ-roma1]{A.~Incicchitti}
\author[kinr]{D.V.~Kasperovych}
\author[kinr]{V.V.~Kobychev}
\author[itep]{S.I.~Konovalov}
\author[lngs]{M.~Laubenstein}
\author[kinr,infn-roma1]{V.M.~Mokina}
\author[kinr,csnsm]{D.V.~Poda}
\author[niic]{V.N.~Shlegel}
\author[kinr]{V.I.~Tretyak}
\author[itep]{V.I.~Umatov}
\author[niic]{Ya.V.~Vasiliev}

\affil[kinr]{Institute for Nuclear Research, 03028 Kyiv, Ukraine}
\affil[itep]{National Research Centre "Kurchatov  Institute of
Theoretical and Experimental Physics, 117218 Moscow, Russia}
 \affil[infn-roma2]{INFN, sezione di Roma ``Tor Vergata'', I-00133 Rome, Italy}
 \affil[univ-roma2]{Dipartimento di Fisica, Universit\`{a} di Roma ``Tor Vergata'', I-00133 Rome, Italy}
 \affil[infn-roma1]{INFN, sezione di Roma, I-00185 Rome, Italy}
 \affil[lngs]{INFN, Laboratori Nazionali del Gran Sasso, I-67100 Assergi (AQ), Italy}
 \affil[kavli]{Kavli Institute for the Physics and Mathematics of the Universe, University of Tokyo, Kashiwa, 277-8583, Japan}
 \affil[csnsm]{CSNSM, Univ. Paris-Sud, CNRS/IN2P3, Universit\'e Paris-Saclay, 91405 Orsay, France}
 \affil[univ-roma1]{Dipartimento di Fisica, Universit\`{a} di Roma ``La Sapienza'', I-00185 Rome, Italy}
 \affil[niic]{Nikolaev Institute of Inorganic Chemistry, 630090 Novosibirsk, Russia}
 \corresp[cor1]{Corresponding author: polischuk@kinr.kiev.ua}

\maketitle

\begin{abstract}
Double beta decay of $^{116}$Cd has been investigated with the
help of radiopure enriched $^{116}$CdWO$_4$ crystal scintillators in the experiment Aurora. The half-life of $^{116}$Cd relatively to the 2$\nu$2$\beta$ decay
of $^{116}$Cd to the ground level of $^{116}$Sn is measured with
the highest up-to-date accuracy as $T_{1/2}$ = [2.69 $\pm$ 0.02
(stat.) $\pm$ 0.14 (syst.)] $\times$ 10$^{19}$ yr. A new improved
limit on the 0$\nu$2$\beta$ decay of $^{116}$Cd to the ground
state of $^{116}$Sn is set as $T_{1/2}\geq 2.4 \times 10^{23}$ yr
at 90\% C.L., that corresponds to the effective Majorana neutrino
mass limit in the range $\langle$$m_\nu$$\rangle$  $\le$ $(1.1-1.6)$ eV, depending on the nuclear matrix elements used in the estimations. New improved
limits on other $2\beta$ processes in $^{116}$Cd (decays with
majoron emission, transitions to excited levels of $^{116}$Sn)
were set at the level of $T_{1/2}\geq 10^{21}-10^{22}$ yr.
\end{abstract}

\section{INTRODUCTION}

Many neutrino experiments give clear evidence for the neutrino
oscillations that can be explained by mixing of neutrinos of different flavors. Taking into account this observation, searches for
neutrinoless double beta ($0\nu2\beta$) decay become very
important experiments to investigate the nature of neutrino
whether being a Dirac or Majorana particle, the lepton number
violation, an absolute scale of neutrino mass and the neutrino
mass hierarchy. Moreover, the $0\nu2\beta$ decay is a powerful
tool to test the Standard Model of particle physics (SM), since
the decay can be mediated by many effects beyond the SM
\cite{Deppisch:2012,Vergados:2016,Delloro:2016,Bilenky:2016}. The
$0\nu2\beta$ decay is still not observed, the most sensitive
experiments give only limits on the decay half-lives for several
nuclei at the level of $\lim T_{1/2}\sim 10^{24}-10^{26}$ yr (see,
e.g., reviews \cite{Cremonesi:2014,Gomes:2015,Sarazin:2015} and
recent results \cite{GERDA,EXO-200,CUORE,NEMO-3,KamLAND-Zen}).
Investigations of the allowed in the SM two neutrino ($2\nu$) mode
of $2\beta$ decay, already detected in several nuclei with the
half-lives in the range of $T_{1/2}\sim10^{18}-10^{24}$ yr (see,
e.g., \cite{Saakyan:2013,Barabash:2015}), give possibility to
examine theoretical calculations of the nuclear matrix elements.
The isotope $^{116}$Cd is one of the most favorable candidates for
$0\nu2\beta$ experiments thanks to the high energy of decay
($Q_{2\beta}=2813.49(13)$ keV \cite{Wang:2017}), promising
estimations of the decay probability
\cite{Rodryguez:2010,Simkovic:2013,Hyvarinen:2015,Barea:2015},
relatively large isotopic abundance ($\delta=7.512(54)\%$
\cite{Meija:2016}) and possibility of enrichment by
ultra-centrifugation in large amount.

\section{EXPERIMENT, RESULTS AND DISCUSSION}

Investigations of 2$\beta$ decay of $^{116}$Cd have been realized
with two $^{116}$CdWO$_4$ crystal scintillators (580
g and 582 g, enriched in $^{116}$Cd to 82\% \cite{Barabash:2011})
in the low background DAMA/R\&D set-up installed deep underground ($\approx$3600~m~w.e.) at the Gran Sasso underground laboratory of
INFN (Italy). The scintillators were fixed inside
polytetrafluoroethylene containers filled with ultra-pure liquid
scintillator, and viewed through low-radioactive quartz light
guides ($\oslash$7 $\times$ 40 cm) by 3 inches low radioactive
photomultiplier tubes (Hamamatsu R6233MOD). The passive shield was
made of high purity copper (10 cm), low radioactive lead (15 cm),
cadmium (1.5 mm) and polyethylene/paraffin (4 to 10 cm) to reduce
the external background. The whole set-up was contained inside a
plexiglas box and continuously flushed by high purity nitrogen gas
to remove environmental radon. An event-by-event data acquisition
system based on a 1 GS/s 8 bit transient digitizer (Acqiris DC270)
recorded the amplitude, the arrival time and the pulse shape of
each events. The energy scale and the energy resolution of the
detectors were checked periodically with $^{22}$Na, $^{60}$Co, $^{133}$Ba, $^{137}$Cs, and $^{228}$Th $\gamma$ sources. The
energy resolution of the $^{116}$CdWO$_4$ detectors for 2615 keV
quanta of $^{208}$Tl was FWHM $\approx$5\%.

\begin{figure}[h]
\centerline{\includegraphics[width=250pt]{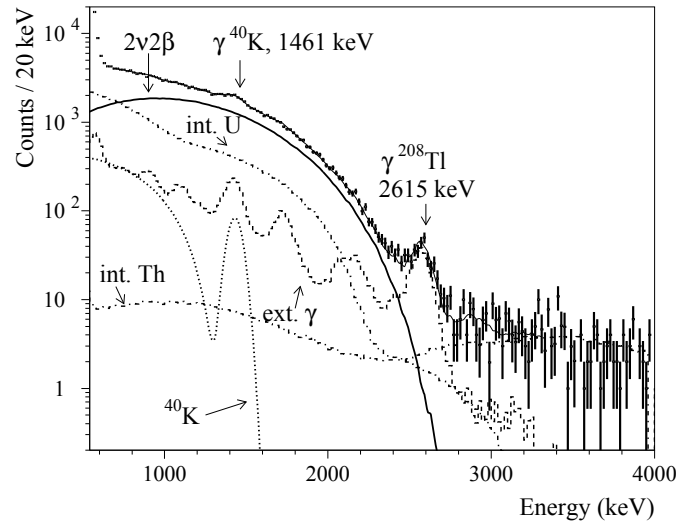}}
 \caption{The energy spectrum of $\gamma$($\beta$) events accumulated over 25037 h with the $^{116}$CdWO$_4$ detectors together with the main
 components of the background model: the $2\nu2\beta$ decay of $^{116}$Cd ("$2\nu2\beta$"), internal contaminations of the $^{116}$CdWO$_4$
 crystals by U/Th, K ("int. U", "int. Th", "$^{40}$K"), and contributions from external $\gamma$ quanta ("ext. $\gamma$").} \label{fig1}
\end{figure}

The pulse profiles of the events were analyzed by using the
optimal filter method \cite{Gatti:1962,Bardelli:2006} to select
$\gamma$ quanta ($\beta$ particles) and $\alpha$ particles. The
pulse-shape discrimination was applied to reduce background and to
estimate (together with the time-amplitude analysis
\cite{Danevich:2001}) the radioactive contamination of the
$^{116}$CdWO$_4$ crystals. The front edge analysis was also used
to reject the fast chains of decays $^{212}$Bi~--~$^{212}$Po from
the $^{232}$Th family. The $^{116}$CdWO$_4$ crystal scintillators
appeared highly radiopure ($\sim$0.02 mBq/kg of $^{228}$Th,
$<0.005$ mBq/kg of $^{226}$Ra, $<0.2$ mBq/kg of $^{40}$K, the
total $\alpha$ activity of U/Th is 2.3(1) mBq/kg). The energy
spectrum of $\gamma$($\beta$) events selected from the data
accumulated over 25037~h with the $^{116}$CdWO$_4$ detectors is
shown in Fig. \ref{fig1}. The spectrum was fitted in the energy
interval 660--3300 keV by the model built from the 2$\nu$2$\beta$
of $^{116}$Cd, the internal contamination by $^{40}$K, $^{232}$Th
and $^{238}$U, and the contribution from external $\gamma$ quanta.
The model functions were Monte Carlo simulated with the EGS4
package \cite{Nel01}, the initial kinematics of the particles
emitted in the decays was given by an event generator DECAY0
\cite{Pon01}. The fit gives the half-life of $^{116}$Cd relatively
to the 2$\nu$2$\beta$ decay to the ground state of $^{116}$Sn as
$T_{1/2}$ = [2.69 $\pm$ 0.02 (stat.) $\pm$ 0.14 (syst.)] $\times$
10$^{19}$ yr. The main contribution to the systematic error comes
from the ambiguity of the radioactive contamination of the
$^{116}$CdWO$_4$ crystals by $^{238}$U, since the $\beta$ spectrum
of $^{234m}$Pa (daughter of $^{238}$U) competes with the
$2\nu2\beta$ spectrum of $^{116}$Cd.

\begin{figure}[h]
    \centerline{\includegraphics[width=200pt]{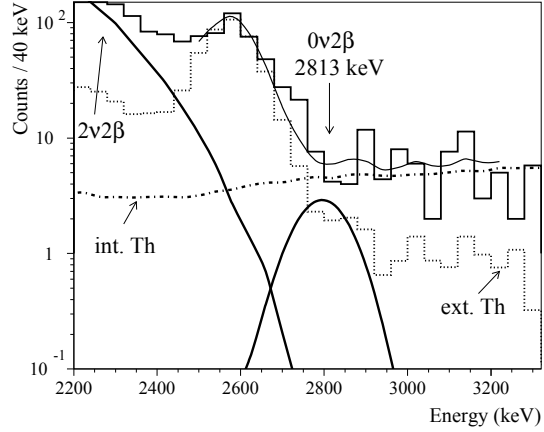}}
    \caption{
The energy spectrum of $\gamma$($\beta$) events accumulated over
28888 h with the $^{116}$CdWO$_4$ detectors in the region of
interest together with the background model: the $2\nu2\beta$
decay of $^{116}$Cd ("$2\nu2\beta$"), the internal contamination
of the $^{116}$CdWO$_4$ crystals by $^{228}$Th ("int. Th"), and
the contribution from external $\gamma$ quanta ("ext. Th").
A peak of the $0\nu2\beta$ decay of $^{116}$Cd excluded at 90\%
C.L. is shown too.} \label{fig2}
\end{figure}

To estimate a limit on the $0\nu2\beta$ decay of $^{116}$Cd we
included in the analysis also the data from the previous stage of
the experiment with a similar background counting rate of $\approx
0.1$ counts/(keV kg yr) in the region of interest (ROI) (see Fig. \ref{fig2}). Furthermore, the background in the ROI was reduced to $\approx
0.07$ counts/(keV kg yr) by selection of the following chain of
decays: $^{212}$Bi ($Q_\alpha=6207$ keV) $\to$ $^{208}$Tl
($Q_\beta$ = 4999 keV, $T_{1/2}$ = 3.053 min). The obtained energy
spectrum was approximated in the energy interval $2.5 - 3.2$ MeV
by the background model constructed from the distributions of the
$0\nu2\beta$ (effect searched for) and $2\nu2\beta$ decays of
$^{116}$Cd, the internal contamination of the crystals by
$^{228}$Th, and the contribution from external $\gamma$ quanta. The fit gives an area of the expected peak $S=-3.2\pm 10.7$
counts, that is no evidence of the effect. In accordance with
\cite{Feldman:1998}, 14.6 counts can be excluded at 90\% confidence
level (C.L.), that leads to the new limit on the $0\nu2\beta$
decay of $^{116}$Cd to the ground state of $^{116}$Sn:
$T_{1/2}\geq 2.4 \times 10^{23}$ yr. The half-life limit
corresponds to the effective neutrino mass limit  $\langle$$m_\nu$$\rangle$  $\le$ $(1.1-1.6)$ eV, obtained by using the recent nuclear matrix elements reported in
\cite{Rodryguez:2010,Simkovic:2013,Hyvarinen:2015,Barea:2015}, the
phase space factor from \cite{Kotila:2012} and the value of the
axial vector coupling constant $g_A=1.27$.

\begin{table}[h]
\caption{The results of the 2$\beta$ experiment with $^{116}$Cd.
The most stringent limits obtained in the previous experiments are
given for comparison at 90\% C.L., except the limits
\cite{Barabash:1990} given at 68\% C.L.}

\label{tab:a}
\tabcolsep7pt
\begin{tabular}{lllll}
\hline
   \tch{1}{l}{b}{Decay \\mode} & \tch{1}{l}{b}{Transition,\\ Level of $^{116}$Sn (keV)}  & \tch{1}{l}{b}{$T_{1/2}$ (yr)\\ }  & \tch{1}{l}{b}{Previous results\\ }   \\
\hline
0$\nu$              & g.s               & $\geq 2.4\times10^{23}$                                           & $\geq 1.7\times10^{23}$ \cite{Danevich:2003} \\
0$\nu$              & $2^{+}$ (1294)    & $\geq 6.2\times10^{22}$                                           & $\geq 2.9\times10^{22}$ \cite{Danevich:2003} \\
0$\nu$              & $0^{+}$ (1757)    & $\geq 6.3\times10^{22}$                                           & $\geq 1.4\times10^{22}$ \cite{Danevich:2003} \\
0$\nu$              & $0^{+}$ (2027)    & $\geq 4.5\times10^{22}$                                           & $\geq 0.6\times10^{22}$ \cite{Danevich:2003} \\
0$\nu$              & $2^{+}$ (2112)    & $\geq 3.6\times10^{22}$                                           & $\geq 1.7\times10^{20}$ \cite{Barabash:1990} \\
0$\nu$              & $2^{+}$ (2225)    & $\geq 4.1\times10^{22}$                                           & $\geq 1.0\times10^{20}$ \cite{Barabash:1990} \\
0$\nu$$M1$          & g.s.              & $\geq 1.1\times10^{22}$                                           & $\geq 8.0\times10^{21}$ \cite{Danevich:2003} \\
0$\nu$$M2$         & g.s.              & $\geq 2.1\times10^{21}$                                           & $\geq 0.8\times10^{21}$ \cite{Danevich:2003} \\
0$\nu$$M3$   & g.s.              & $\geq 0.9\times10^{21}$                                           & $\geq 1.7\times10^{21}$ \cite{Danevich:2003} \\
2$\nu$              & g.s               & $[2.69\pm 0.02$(stat.)$\pm0.14$ (syst.)]$\times 10^{19}$  & see Table 2 in \cite{Polischuk:2015} and work \cite{Arnold:2017} \\
2$\nu$              & $2^{+}$ (1294)    & $\geq 0.9\times10^{21}$                                           & $\geq 2.3\times10^{21}$ \cite{Piepke:1994} \\
2$\nu$              & $0^{+}$ (1757)    & $\geq 1.0\times10^{21}$                                           & $\geq 2.0\times10^{21}$ \cite{Piepke:1994} \\
2$\nu$              & $0^{+}$ (2027)    & $\geq 1.1\times10^{21}$                                           & $\geq 2.0\times10^{21}$ \cite{Piepke:1994} \\
2$\nu$              & $2^{+}$ (2112)    & $\geq 2.3\times10^{21}$                                           & $\geq 1.7\times10^{20}$ \cite{Barabash:1990} \\
2$\nu$              & $2^{+}$ (2225)    & $\geq 2.5\times10^{21}$                                           & $\geq 1.0\times10^{20}$ \cite{Barabash:1990} \\
\hline
\end{tabular}
\end{table}

Limits on other
2$\beta$ processes in $^{116}$Cd were derived in a similar way.
The results are reported in Table 1.

\section{CONCLUSIONS}
The Aurora experiment to investigate 2$\beta$ processes in
$^{116}$Cd with enriched $^{116}$CdWO$_4$ scintillators (1.16~kg) is
finished after 33737 h (total time) of measurements at the Gran Sasso
underground laboratory of INFN (Italy). The half-life of
$^{116}$Cd relatively to the $2\nu2\beta$ decay to the ground
state of $^{116}$Sn is measured with the highest up-to-date
accuracy: $T_{1/2}$ = [2.69 $\pm$ 0.02~(stat.) $\pm$ 0.14~(syst.)]
$\times$ 10$^{19}$ yr. A new $0\nu2\beta$ half-life limit is set
as $T_{1/2}$ $\ge$ 2.4$\times$ 10$^{23}$ yr at 90\% C.L., that
corresponds to the effective Majorana neutrino mass limits
$\langle$$m_\nu$$\rangle$  $\le$ $(1.1-1.6)$ eV, depending on the nuclear matrix elements used in the analysis. Other $2\beta$ processes in
$^{116}$Cd are restricted at the level of
$T_{1/2}>(0.1-6.3)\times 10^{22}$ yr.


\nocite{*}
\begin{thebibliography}{99}
 \bibitem{Deppisch:2012} F.F.~Deppisch, M.~Hirsch, H. P\"{a}s, {\it J. Phys. G} {\bf39}, 124007 (2012).
 \bibitem{Vergados:2016} J.D.~Vergados, H.~Ejiri, F.~Simkovic, {\it Int. J. Mod. Phys. E} {\bf25}, 1630007 (2016).
 \bibitem{Delloro:2016} S.~Dell'Oro, S.~Marcocci, M.~Viel, F.~Vissani, {\it AHEP} {\bf2015}, 2162659 (2016).
 \bibitem{Bilenky:2016} S.M.~Bilenky, C.~Giunti, {\it Int. J. Mod. Phys. A} {\bf30}, 1530001 (2015).
 \bibitem{Cremonesi:2014} O.~Cremonesi, M.~Pavan, {\it AHEP} {\bf2014}, 951432 (2014).
 \bibitem{Gomes:2015} J.J.~G$\mathrm{\acute{o}}$mez-Cadenas, J.~Mart$\mathrm{\acute{i}}$n-Albo, {\it Proc. of Sci.} {\bf(GSSI14)}, 004 (2015).
 \bibitem{Sarazin:2015} X.~Sarazin, {\it J. Phys.: Conf. Ser.} {\bf593}, 012006 (2015).
 \bibitem{GERDA} M.~Agostini et al., {\it Nature} {\bf544}, 47 (2017).
 \bibitem{EXO-200} J.B.~Albert et al., {\it Nature} {\bf510}, 229 (2014).
 \bibitem{CUORE} K. Alfonso et al., {\it Phys. Rev. Lett.} {\bf115}, 102502 (2015).
 \bibitem{NEMO-3} R. Arnold et al., {\it Phys. Rev. D} {\bf92}, 072011 (2015).
 \bibitem{KamLAND-Zen} A.~Gando et al., {\it Phys. Rev. Lett.}, {\bf117}, 082503 (2017).
 \bibitem{Saakyan:2013} R.~Saakyan, {\it Annu. Rev. Nucl. Part. Sci.} {\bf 63}, 503 (2013).
 \bibitem{Barabash:2015} A.S.~Barabash, {\it Nucl. Phys. A} {\bf 935}, 52 (2015).
 \bibitem{Wang:2017} M.~Wang et al., {\it Chin. Phys. C} {\bf 41}, 030003 (2017).
 \bibitem{Rodryguez:2010} T.R.~Rodr\'iguez, G.~Mart\'inez-Pinedo, {\it Phys. Rev. Lett.} {\bf105}, 252503 (2010).
 \bibitem{Simkovic:2013} F.~\v{S}imkovic, V.~Rodin, A.~Faessler, P.~Vogel, {\it Phys. Rev. C} {\bf87}, 045501 (2013).
 \bibitem{Hyvarinen:2015} J.~Hyv$\mathrm{\ddot{a}}$rinen, J.~Suhonen, {\it Phys. Rev. C} {\bf91}, 024613 (2015).
 \bibitem{Barea:2015} J.~Barea, J.~Kotila, F.~Iachello, {\it Phys. Rev. C} {\bf91}, 034304 (2015).
 \bibitem{Meija:2016} J.~Meija et al., {\it Pure Appl. Chem.} {\bf 88}, 293 (2016).
 \bibitem{Barabash:2011} A.S.~Barabash et al., {\it JINST} {\bf 6}, P08011 (2011).
 \bibitem{Gatti:1962} E.~Gatti, F.~De~Martini, Nuclear Electronics II, Proceedings of the Conference on Nuclear Electronics. V. II, International Atomic Energy Agency, Vienna, 1962.
 \bibitem{Bardelli:2006} L.~Bardelli et al., {\it Nucl. Instr. Meth. A} {\bf 569}, 743 (2006).
 \bibitem{Danevich:2001} F.A.~Danevich et al., {\it Nucl. Phys. A} {\bf 694}, 375 (2001).
 \bibitem{Nel01} W.R.~Nelson et al., SLAC-Report-265, Stanford, 1985.
 \bibitem{Pon01} O.A.~Ponkratenko et al., {\it Phys. At. Nucl.} {\bf63}, 1282 (2000); V.I. Tretyak, to be published.
 \bibitem{Feldman:1998} G.J.~Feldman, R.D.~Cousins, {\it Phys. Rev. D} {\bf57}, 3873 (1998).
 \bibitem{Kotila:2012} J.~Kotila, F.~Iachello, {\it Phys. Rev. C} {\bf85}, 034316 (2012).
 \bibitem{Danevich:2003}  F.A.~Danevich et al., {\it Phys. Rev. C} {\bf67}, 014310 (2003).
 \bibitem{Barabash:1990} A.S.~Barabash, A.V.~Kopylov, V.I.~Cherehovsky, {\it Phys. Lett. B} {\bf249}, 186 (1990).
 \bibitem{Polischuk:2015} O.G.~Polischuk et al., {\it AIP Conf. Proc.} {\bf1686}, 020017 (2015).
 \bibitem{Arnold:2017} R.~Arnold et al., {\it Phys. Rev. D} {\bf95}, 012007 (2017).
 \bibitem{Piepke:1994} A.~Piepke et al., {\it Nucl. Phys. A} {\bf577}, 493 (1994).
\end{thebibliography}
\end{document}